\documentstyle[aps,epsfig,prl]{revtex}

\begin{document}

\title{On the structure and stability of germanium nanoparticles}


\author{Laurent Pizzagalli\footnote{Permanent address: Laboratoire de 
M\'etallurgie Physique, SP2MI, BP 30179, F-86960 Futuroscope cedex, France.}, 
Giulia Galli, John E.~Klepeis and Fran\c{c}ois Gygi }

\address{Lawrence Livermore National Laboratory, P.O. Box 808, L-415, Livermore CA 94551
\vspace{4ex}
\parbox{14.3cm}
{\rm
\hspace{1.5em}
\begin{abstract}
In order to control and tailor the properties of nanodots, it is essential to
separate the effects of quantum confinement  from those due to the surface,
and to gain insight into the influence of 
preparation conditions on the dot physical properties.
We address these issues for the case of small Ge clusters (1-3 nm),
using a combination of empirical and first-principles molecular dynamics techniques.
Our results show that over a wide temperature range the diamond structure
is more stable than tetragonal, ST12-like structures for clusters containing
more than 50 atoms; however, the magnitude of the energy
difference between the two geometries is strongly dependent on the surface
properties. 
Based on our structural data, we propose a mechanism which may be 
responsible for the formation of metastable
ST12 clusters in vapor
deposition experiments, by cold quenching of amorphous nanoparticles with unsaturated,
reconstructed surfaces.
\end{abstract}
}}
\maketitle
 In semiconductor nanoparticles quantum confinement leads to
an increase of the optical gap compared to the bulk value and thus 
opens new possibilities for controlling photoluminescence effects,
with narrow emission spectra tunable over a wide range of wavelengths\cite{rev1,rev2}. 
These properties make semiconductor dots attractive for many applications
including photovoltaics, lasers and infrared dyes.
Furthermore, their brightness, low toxicity 
and the ability to use a single
excitation wavelength make them good alternatives to organic dyes
for biological labelling, but their low water 
solubility has limited biological applications. However, recent experiments 
have shown that using specific coatings, the surface of selected semiconductor
nanodots can be tailored to enhance the chemical interaction with a biological sample
and the water solubility\cite{dot1,dot2}.

 Understanding the influence of surface reconstruction and passivation on
the ground state properties of semiconductor nanodots is a key
prerequisite not only in designing biological applications,
but also for 
controlling deposition of nanoparticles on surfaces and aggregation of multiple
dots into new structures.
In order to tailor the properties of nanodots, it is important to
separate the effects of quantum confinement  from those due to the surface,
and to gain insight into the mechanism by which
preparation conditions can influence the dot atomic structure and thus its
optical properties.

 Here we address these issues for the case of small Ge dots (1-3 nm),
whose atomic structure is the most controversial amongst those
of group IV and II-VI semiconductors.
While some preparation techniques, including  chemical 
methods\cite{Car96CM,Min96APL,Mig98APL,Tay98CM,Guh99NIMPR},  
yield diamond-like Ge dots irrespective of size,
several experiments\cite{Sai79JCG,Kan92APL,Jia94APL} suggest a structural 
transition, as the dot diameter becomes smaller than 4-5 nm.
In particular, some experiments\cite{Sat95APL,Sat98APL} using vapor deposition techniques indicate
a change from a cubic diamond (DIA) to a tetragonal structure, possibly ST12,
in contrast to  the behavior found for Si\cite{Tol96PRL,Din96PRB,Buu98PRL} 
and other II-VI dots\cite{Tol94SCI,Che97SCI}.
In the bulk, the ST12 phase is only obtained from high-pressure  
experiments \cite{Bun63SCI}. 
The ST12 gap is direct and 
about 1.5~eV, as opposed to an indirect 
gap of 0.7~eV in the cubic phase.
The relationship between a possible structural transition as a function 
of nanoparticle size and cluster preparation conditions, as well as the influence of a  
structural transition on the dot optical properties are as yet unknown.

 Using first principles calculations we have studied the
structure and stability of Ge clusters with diameters smaller
than approximately 2.5 nm. We have considered both
cubic diamond and tetragonal ST12 structures with 
H-terminated  and bare reconstructed
surfaces. Summarizing our results, we found
that over a wide temperature range, the DIA
structure is more stable than ST12-like structures for all clusters
with more than 40 atoms (about 1 nm in diameter). 
The energy difference between the two geometries
monotonically increases  as a function of size, for both
H-terminated and bare reconstructed surfaces.
However the magnitude of this energy difference
is strongly dependent on the 
reconstruction and on H-passivation.
Our calculations also suggest that surface effects may be responsible for
the formation of metastable ST12 clusters from amorphous nanoparticles
in vapor deposition experiments: when the dot diameter becomes smaller than $\simeq$ 2-3 nm, 
a structural transition from 
amorphous to ST12-like 
nanodots may be driven by the pressure on the inner core of the dot arising
from the reconstructed surface. 

In our calculations, Ge nanocrystallites were represented by 
free-standing clusters \cite{Ge6note}
in a large supercell \cite{sizecell}. 
The total energy of the dots was computed 
using Density Functional Theory (DFT) in the Local
Density Approximation (LDA), using iterative optimization techniques\cite{Car85PRL}.
The electronic wave functions 
were expanded in plane waves, with an energy cutoff of 11~Ry,
and non-local pseudopotentials were used to 
represent the interaction between the electrons and ionic cores\cite{Ham89PRB}. 
We considered nanoclusters with spherical shapes, 
with a number of Ge atoms ranging from 28 to 300\cite{Ge3note}.
 In all cases we considered sizes which allowed us to use the same number of
Ge and H atoms for both DIA and ST12 structures, in order to have direct total
energy comparisons \cite{Ge4note}.

 The energy difference between  DIA and ST12 Ge clusters with H-saturated surfaces 
is plotted as a function of cluster size in Fig.~1 (dotted line).
Our results show that H-passivated dots with a  DIA-like  structure are more stable 
than those with a ST12-like structure for all sizes, with the
energy difference increasing as a function of the nanoparticle diameter. 
The average volume  per atom ($V$) of the cluster, shown in Fig.~2, 
is reduced in comparison to the volume per atom ($V_0$) corresponding to the
bulk first neighbor distances for both DIA and ST12 geometries; 
for example, the ratio $V/V_0$  for DIA (ST12) dots varies from 0.97 (0.96) to 0.99 (0.98) 
when going from a 45 to a 145 atom cluster. Over the same range of sizes, the 
effective pressure acting on the
cluster core, as evaluated using bulk moduli data, decreases from about 2 to 1 GPa.
The magnitude of the DIA-ST12 
energy difference, as well as the contraction of the average cluster atomic volume are 
significantly modified in the presence of unsaturated, reconstructed surfaces, 
as described  below.

 In order to find reconstructed geometries for the cluster surfaces we used a 
combination of empirical and DFT-LDA techniques.
Formation of facets is expected for clusters larger than 5~nm, 
but facets are unlikely in spherical nanoclusters with 1-5 nm diameters\cite{rev2}.
In addition clusters prepared by deposition on surfaces usually
exhibit disordered, defected interfaces\cite{rev2}. 
We therefore chose to determine surface 
reconstructed geometries using an annealing procedure. 
First, using  molecular dynamics with a Tersoff potential \cite{Ter89PRB}, 
we melted the cluster surface by heating it up to 2000~K for 0.08~ns; 
the temperature was then slowly decreased 
to zero over 0.2-0.5~ns. During this phase of the calculation,
the crystalline core of the nanoclusters was kept frozen, and the 
shape conserved by confining the system in a spherical cavity. 
The annealing/quenching series was repeated 
2-4 times . The final structure was fully 
relaxed within DFT-LDA\cite{Car85PRL}, all of the atoms being allowed to move.
During these relaxations, energy gains varied from 310 (275) meV/atom
for 95 atom clusters to 88 (81) meV/atom for 300 atom clusters,
for dots with a DIA (ST12)-like core structure.
Fig.~3 shows the surface structure and the crystalline core for a 
selected cluster (Ge$_{\rm 190}$).

 As shown in Fig.~1 (dotted versus dashed lines), we observed a strong reduction 
of the energy difference [$\Delta E(N)$] between 
DIA and ST12 clusters when the H-passivated surface is replaced by one which is 
unsaturated and reconstructed. 
In the absence of H atoms, and given the spherical shape of the clusters,
the computed values of $\Delta E(N)$ can be fitted by separating surface
and bulk contributions: 
$\Delta E(N)=N\Delta\epsilon+(36\pi)^{\frac{1}{3}}N^{\frac{2}{3}}\Delta\gamma $.
Here $\Delta\epsilon = -87$ meV/at  and $\Delta\gamma=+59$~meV/at are the volume 
and surface contributions to the energy difference $\Delta E(N)$, respectively.
The value of $\Delta\epsilon$ is smaller than the calculated energy difference
between solid DIA and ST12 (for the bulk energy difference we find 130 meV/atom, 
in good agreement  with previous calculations\cite{Muj93PRB,Cra94PRB}).  
As indicated by the value of $\Delta\gamma=+59$~meV/at,  surface energy
is smaller for ST12 clusters  and its sign is opposite to that of the
volume contribution to $\Delta E(N)$. This is largely responsible for the reduction of
$\Delta E$ in dots with reconstructed surfaces, compared to H-saturated clusters.
Our results indicate that the ST12 structure should be more stable than DIA for $N\leq40$; 
however for such small sizes,  Ge clusters are not expected to 
exhibit bulk-like geometries, but rather to form complex, non spherical shapes.

 As mentioned above, the average atomic volume of a dot with a bare
reconstructed surface is reduced in comparison to that of a 
cluster with a hydrogenated surface \cite{calcpress}. As plotted in Fig.~2, the ratio V/V$_0$ 
is 0.93 (0.947) for ST12 (DIA) Ge$_{190}$ clusters. The volume reduction remains 
strong for our biggest Ge$_{300}$ clusters, with 0.954 (0.967) for ST12 (DIA).   
These last values are to be compared, e.g.,
with the value of 0.982 (0.988) for ST12 (DIA) in the case of the 
smaller Ge$_{145}$H$_{108}$ dot, which has roughly the
same number of atoms belonging to the crystalline dot core as Ge$_{300}$.
The average volume reduction of the cluster amounts to an effective pressure on 
the crystalline core of about 4 and 2.3 GPa, for
Ge$_{190}$ and Ge$_{300}$, respectively.
The effective pressure is slightly higher for ST12 than for DIA geometries (see Fig.~2).

 An analysis of the reconstructed surfaces reveals disordered structures in all cases, 
as expected from 
a fast quench from a liquid state. For the larger clusters, the bond 
angles range approximately from 63~$^\circ$ to 144~$^\circ$, and 
the average bond length is close to the 
calculated first neighbor-distances in amorphous Ge  (2.46~\AA), i.e. 2~\%\ larger than in
the crystalline DIA structure.
In general, we observed a strong reduction of the undercoordinated surface atoms after 
reconstruction, due to atomic dimerization. 
This effect is stronger for ST12 than for diamond, with ST12 reconstructed 
nanoclusters exhibiting approximately 20\%\ fewer dangling bonds.
This circumstance is due to the smaller size of the ST12 crystalline core
and to the broader distribution of bond angles in
the bulk ST12 structure, both of which provide
greater freedom in the rearrangement of surface atoms.
For one cluster size, we studied the effect on the dot stability
of H-passivating the  dangling bonds of its
reconstructed surface. We observed an increase of $\Delta E$,
the DIA structure being favored over ST12 
(the full relaxation of selected hydrogenated reconstructed surfaces 
were carried out within DFT-LDA). 

 As a final step in our study of the stability of Ge nanodots, we  estimated the effect of temperature on 
$\Delta E(N)$, for clusters with reconstructed surfaces, by computing free energy differences in the
harmonic approximation. 
The vibrational free energy 
$F_{vib}=\sum^{3N-6}_{i=1}\left[\frac{\hbar\omega_i}{2}+
k_BT\ln\left(1-\exp(-\frac{\hbar\omega_i}{k_BT})\right)\right]$ 
was determined by computing the  vibrational frequencies $\omega_i$ using the Tersoff potential. 
Although not as accurate as total energy differences obtained within DFT-LDA, $F_{vib}$  can be
used to estimate finite temperature effects  as a function of size \cite{Ge2note}.
Our results, shown in Fig.~1 by open dots, indicate that 
energy differences between DIA and ST12 are slightly reduced. However 
temperature effects  do not invert the relative stability of the two structures. 
For examples, at $N=145$
a temperature greater than 1180~K, i.e. close to the melting point 
of Ge, would be required for the 
reconstructed ST12 cluster to be more stable than DIA.

 Our total energy calculations  have shown that 
cubic diamond is the most stable structure of both H-terminated and bare
reconstructed Ge clusters in the 1-3~nm size range, despite the importance of surface 
reconstruction effects. 
Ge dots with the ST12 structure are metastable and it is interesting  
to investigate whether there exist experimental conditions which might give rise to metastable
ST12 clusters. For example, in vapor deposition or sputtering experiments\cite{Bla98JAC},
amorphous Ge nanoparticles are initially present and annealing treatments
are usually required for crystallization to occur. It is therefore
relevant to understand whether metastable ST12 nanoparticles can be quenched 
from amorphous dots.

 Based on our calculations, the cores
of ST12 and DIA dots with both H-terminated and reconstructed surfaces are compressed.
The effective pressure on the dot cores is much larger in the presence of 
reconstructed surfaces.
These results suggest that pressure effects may play a role in quenching
metastable  ST12 clusters from amorphous nanoparticles.
In order to address this issue, we have first investigated  the 
amorphous (a-Ge) to ST12 transition in bulk Ge. Fig.~4 shows the total energy of diamond, ST12 and
a-Ge as a function of volume, at T=0, as obtained from our calculations.
Both a-Ge and ST12 are metastable, with the amorphous phase being slightly lower in energy than 
the ST12 crystal.  A pressure P$_t$ of 1.5~GPa
is required to induce an a-Ge to ST12 transition. Whether such a transition actually 
occurs depends
on the height of the barrier between the two structures and on the
temperature. We have not attempted to compute the a-Ge to ST12 energy barrier;
however, phenomenological arguments suggest that it should be lower than that 
between  a-Ge and cubic diamond,  at temperatures typical of, e.g. dot deposition experiments.
Indeed, ST12 is a weakly ordered crystal, which has 
been used to model a-Ge: it has 12 atoms per unit cell and a space group with few symmetry 
operations \cite{Ge7note}.
Most importantly, unlike diamond, the ST12 crystalline network  exhibits 7- and 5-fold atom rings, similar to  a-Ge.
It is therefore conceivable that a transition between 
a-Ge and ST12 be possible at relatively low T, when
a-Ge is under a pressure of 1.5 GPa or higher. 
Similar considerations for the case of nanodots suggest that 
the pressure exerted by reconstructed  surfaces on amorphous nanoparticle cores  
initially present in vapor deposition experiments may be large enough to induce a
transition from amorphous to ST12 metastable nanoparticles.
An extrapolation of the calculated effective pressures on dot cores (see Fig.~2) suggest that
for dots with bare reconstructed surfaces and a diameter smaller than 2.5-3~nm, 
the pressure on the crystalline core  is larger than the pressure required
in the bulk to induce an a-Ge to ST12 transition (2.5-3~GPa, as compared to 1.5 GPa).
Therefore for dots with diameters smaller than 2.5-3.0 nm prepared in vapor deposition 
experiments\cite{Sat95APL,Sat98APL},
an a-Ge to ST12 transition induced by an effective {\em surface pressure} may be possible.
On the contrary, the pressure exerted on the core of H-passivate clusters  is equal to
or smaller than the bulk transition pressure even for clusters with 70-100~atoms (i.e. with
a diameter less than 1.5~nm).

The LDA used in our work does not permit quantitative evaluations
of optical gaps for Ge dots. However, it is interesting to note that the difference 
between the energy of the highest occupied orbital (HOMO) and lowest unoccupied orbital (LUMO)
as obtained for H-passivated clusters is larger for DIA-like geometries
than for ST12-like geometries\cite{Ge8note}.
While the HOMO position in energy is similar in DIA and ST12 clusters, the LUMO
position is higher for DIA than for ST12 clusters.
If the same trend was to be confirmed for quasi-particle energies, then a measure of the
optical gap of small Ge dots could be a way to discriminate between DIA and ST12
geometries. Work is in progress to go beyond the LDA and estimate optical gaps.

In conclusion, we have shown that Ge clusters with the diamond structure
are more stable than tetragonal ST12 dots over a wide temperature range, 
irrespective of the cluster size, for dot diameters larger than $\simeq$ 1.5 nm.
We have proposed a mechanism which may be responsible for the formation of metastable 
ST12 clusters in vapor
deposition experiments\cite{Sat95APL,Sat98APL}, by cold quenching of amorphous nanoparticles with unsaturated, 
reconstructed surfaces.
The pressure exerted on the nanoparticle core by the surface can induce an amorphous to 
ST12 transition, for clusters
with diameters smaller than 2.5-3.0 nm. This may explain why different types of 
structures are seen
in experiments using chemical preparation methods\cite{Car96CM,Mig98APL,Tay98CM} versus 
physical vapor deposition methods\cite{Sat95APL,Sat98APL}.
According to our calculations, chemical methods should always yield diamond structures, 
consistent with the results of H.~W.~Lee et al.\cite{Leecomm}.
Our study indicates that quantum confinement as well as surface effects are both
key features
in understanding the physical properties of small semiconductor dots. 
By tuning the surface properties with, for example, a 
particular choice of surfactant or by otherwise controlling the surface reconstruction, 
the pressure exerted on the dot core can be modified and 
used to tailor the atomic structure of the dot and indirectly the electronic properties.

Useful discussions with L.~Terminello, C.~Bostedt, A.~Van~Buuren, H.~Lee, 
S.~Bastea and E.~Schwegler are gratefully
acknowledged.
This work was performed under the auspices of the U.S. Department of
Energy by University of California Lawrence Livermore National
Laboratory, Office of Basic Energy Sciences, Division of Materials Science,
under contract No. W-7405-Eng-48.
%
\begin{figure}
{\par\centering \resizebox*{7.5cm}{!}{
\includegraphics{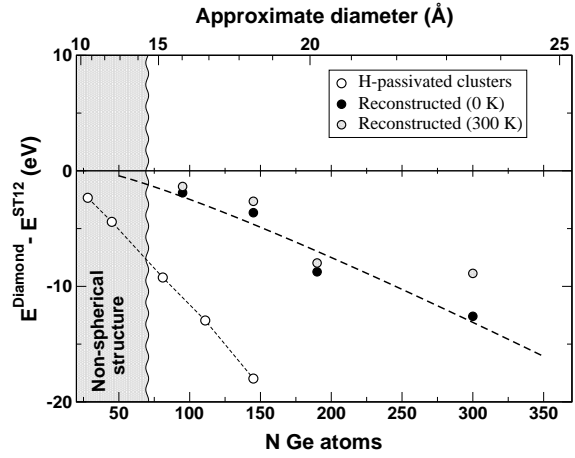}}\par}
\vskip 2.0cm
\caption{Energy differences between Ge dots with diamond and ST12 structures as a 
function of the number of atoms and the approximate dot diameter (top horizontal axis).
We show results for clusters with H-passivated, non-reconstructed surfaces  at 0~K,
and for clusters with bare reconstructed surfaces at 0~K (black circles) 
and 300~K (grey circles).
The approximate boundary between clusters with complex, non-spherical shapes and
dots with crystalline like geometries has been drawn to guide the eye, based on the data of
Hunter et al \protect\cite{Hun94PRL}.}
\end{figure}
%
\begin{figure}
{\par\centering \resizebox*{7.5cm}{!}{
\includegraphics{new_fig2.eps}} \par}
\vskip 2.0cm
\caption{
\protect
Ratio between the average atomic volume in Ge clusters ($V$) and the bulk equilibrium 
atomic volume ($V_0$),
as calculated within the LDA (upper panel), and the
corresponding internal pressure (lower panel) as a function of the cluster 
size.
Filled (empty) circles represent 
H-passivated diamond (ST12) nanoclusters;
filled (empty) squares represent diamond (ST12) dots with bare reconstructed
surfaces. In the lower panel the horizontal axis indicates the
transition pressure between amorphous and ST12 crystalline germanium in the bulk.}
\end{figure}
%
\begin{figure}
{\par\centering \resizebox*{4.5cm}{!}{
\includegraphics{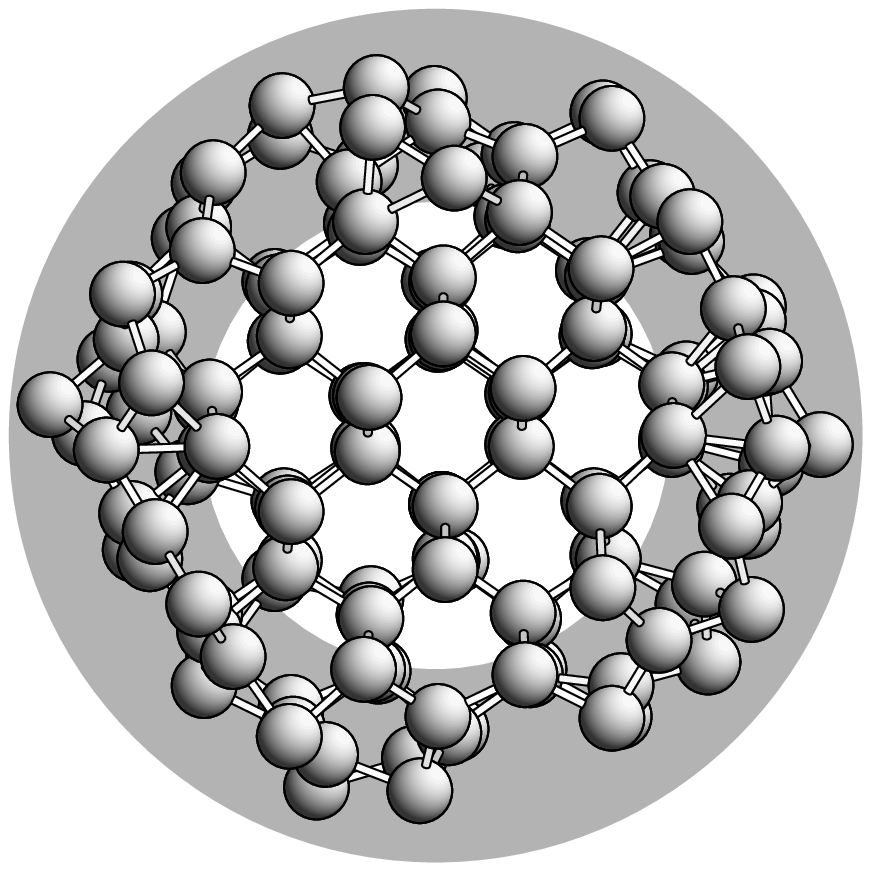}} \par}

\caption{
\protect
Cross-section view of  Ge$_{190}$ with a crystalline diamond like core,
indicated by the white area. The grey area indicates 
the cluster disordered surface.}
\end{figure}
%
\begin{figure}
{\par\centering \resizebox*{7.5cm}{!}{{\rotatebox{-90}{
\includegraphics{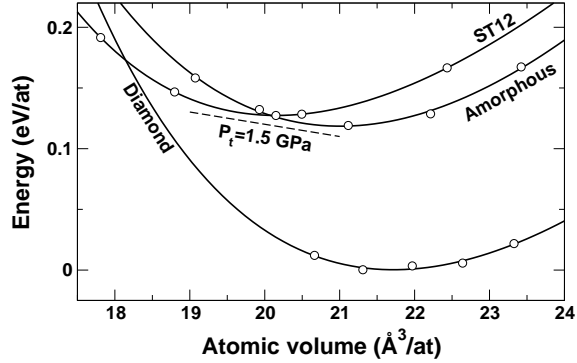}}}} \par}
\vskip 2.0cm
\caption{
Total energy per atom as a function of the atomic volume for bulk Ge in the
diamond, ST12 and amorphous (a-Ge) structures. The Murnaghan equation of
state fits are shown as solid lines. 
The transition pressure (P$_t$) between a-Ge and ST12 is
equal to the slope of the common tangent to the ST12 and a-Ge total 
equation of state curves.
The diamond curve has been calculated by varying the lattice parameter
of a cubic cell containing 216 Ge atoms. This corresponds to a sampling of 5~k-points 
in the Irreducible Brillouin Zone (IBZ). The structural parameters of the ST12 structure have 
been optimized by relaxing simultaneously the ionic positions and lattice 
parameters of a 96 atom
supercell. Then the total energy at the minimum has been recomputed using a 324 atom cell, 
thus allowing
for a better k-point sampling (13 k-points in the IBZ).
Amorphous samples with 144 atom have
been prepared first by using a Tersoff potential, with thermalization to 2000~K 
and subsequent slow quenches. 
The final relaxations have been carried out within DFT-LDA and the atomic positions have 
been optimized at different densities, with constant pressure ab-initio MD runs.
Finally, the energy difference between diamond and a-Ge was adjusted to the 
experimental enthalpy difference between the two phases\protect\cite{Che69JAP}. 
}
\end{figure}


\begin{thebibliography}{10}
\bibitem{sizecell}
We used a cubic supercell with a lateral dimension $a=60$~a.u. For the largest clusters
we have verified that the charge density is negligible at the 
supercell boundary.
\bibitem{Ge4note}
For diameter smaller than 5~nm, nanoparticles are commonly observed with
  spherical shapes \cite{Car96CM,Min96APL,Guh99NIMPR,Kan92APL}. However, it has
  been shown that nanoclusters with less than 70~Ge atoms ($\sim1.4$~nm)
  exhibits a non-spherical shape \cite{Hun94PRL}. Since the focus of our
  investigation is on the properties of Ge dots in the range 1-5~nm, we
  considered only spherical clusters in this work.
\bibitem{calcpress}
The atomic volume inside the clusters is calculated by comparing the relaxed configuration with a 
ideal reference system (built by using the bulk lattice parameter calculated by DFT-LDA). The effective pressure 
corresponding to the atomic volume reduction is then obtained from the results of the constant-pressure calculations 
detailed in the caption of the Figure~4.
\bibitem{Ge2note}
The vibrational properties of Ge as given by the Tersoff potential are in a
  qualitative agreement with experiments and DFT calculations \cite{Mor98JJAP}.
  For instance, we calculated a Ge$_2$ stretching mode frequency of
  251~cm$^{-1}$ with the Tersoff potential, close to 245~cm$^{-1}$ as
  determined in DFT-LDA.
\bibitem{Ge7note}
The space group of ST12, P$4_32_12$, has 8 symmetry operations, to compare with the 
192 symmetry operations of the space group Fd3m of the cubic diamond structure.
\bibitem{Ge8note}
We estimate that the LDA HOMO-LUMO gap of H-passivated Ge nanodots becomes larger for ST12-like than DIA-like
structures when the dot diameter is greater than 3~nm.
\end{thebibliography}
\end{document}